\documentclass[prl, 10pt, twocolumn, superscriptaddress]{revtex4-1}

\usepackage{amsmath, amssymb}
\usepackage{graphicx,color,xfrac,bm,stmaryrd,trimclip,array,wasysym}
\definecolor{lightblue}{rgb}{0.17,0.39,1}
\usepackage[normalem]{ulem}
\usepackage[utf8x]{inputenc}
\usepackage{titlesec}
\titlespacing*{\section}{0pt}{*1.5}{*0.5}

\usepackage[colorlinks=true, allcolors=lightblue, linktocpage=true]{hyperref}
\definecolor{lightblue}{rgb}{0.17,0.39,1}
\definecolor{lightgreen}{rgb}{0.67,0.81,0.08}
\definecolor{lightred}{rgb}{1,0.05,0.52}
\hypersetup{linkcolor=lightblue, citecolor=lightblue, filecolor=black, urlcolor=lightblue}

\begin{document}

\title{Emergent low-energy thermodynamics and universal $B$/$T$ scaling in $\alpha$-RuCl$_3$}

\author{A.~Khansili}
\email[Email: ]{akash.khansili@gmail.com}
\affiliation{Institute of Science and Technology Austria, Am Campus 1, 3400 Klosterneuburg, Austria}				
\affiliation{Department of Physics, Stockholm University, SE-106 91 Stockholm, Sweden}

\author{B.~J.~Ramshaw}
\affiliation{Laboratory of Atomic and Solid State Physics, Cornell University, Ithaca, New York 14853, USA}
\affiliation{Canadian Institute for Advanced Research, Toronto, Ontario, Canada}

\author{A.~Bangura}
\affiliation{National High Magnetic Field Laboratory, Florida State University, Tallahassee, FL 32310, USA}

\author{H.-Y.~Kee}
\affiliation{Department of Physics, University of Toronto, Ontario, Canada M5S 1A7}
\affiliation{Canadian Institute for Advanced Research, CIFAR Program in Quantum Materials, Toronto, Ontario, Canada, M5G 1M1}

\author{R.~D.~McDonald}
\affiliation{Los Alamos National Laboratory, Los Alamos, New Mexico 87545, USA}

\author{F.~Assaad}
\affiliation{Institute for Theoretical Physics, Julius-Maximilians-Universität Würzburg, 97074 Würzburg, Germany}
\affiliation{Würzburg-Dresden Cluster of Excellence ctd.qmat, Julius-Maximilians-Universität Würzburg, 97074 Würzburg, Germany}

\author{A.~Rydh}
\affiliation{Department of Physics, Stockholm University, SE-106 91 Stockholm, Sweden}

\author{A.~Shekhter}
\email{Deceased}
\affiliation{Los Alamos National Laboratory, Los Alamos, New Mexico 87545, USA}

\author{K.~A.~Modic}
\affiliation{Institute of Science and Technology Austria, Am Campus 1, 3400 Klosterneuburg, Austria}


\begin{abstract}
Quantum spin liquids are exotic states of matter in which strongly interacting spins remain disordered down to zero temperature and host fractionalized excitations. This has motivated an extensive search for their realization in materials. Among candidate systems, $\alpha$-RuCl$_3$ has drawn particular attention as a platform for the quantum spin liquid proposed by Kitaev, in which the low-energy excitations in applied magnetic fields are interpreted as itinerant Majorana fermions and local flux excitations. However, conflicting experimental signatures have cast doubt on whether a Kitaev quantum spin liquid-like state is really the right starting point to interpret the low-energy states in $\alpha$-RuCl$_3$, motivating a closer look at the nature of these excitations. Here we report an extensive high field study of specific heat up to 35\,T on $\alpha$-RuCl$_3$. We find that the low-energy thermodynamics exhibit universal $B/T$ scaling with a single effective energy scale proportional to the applied field as long as the magnetic field is not close to the $c$-axis. This behavior persists continuously from the antiferromagnetic phase boundary to the highest fields, suggesting that the perturbative Kitaev limit, with its Majorana and flux excitations, may not be an appropriate starting point for describing the low-energy thermodynamics of $\alpha$-RuCl$_3$ for fields away from the $c$-axis.
\end{abstract}


\date{\today}
\maketitle
\setlength{\parindent}{0.5cm}
\setlength{\parskip}{0.2cm}

\section{Introduction}

\begin{figure*}[t!!]
\includegraphics[width=0.83\textwidth]{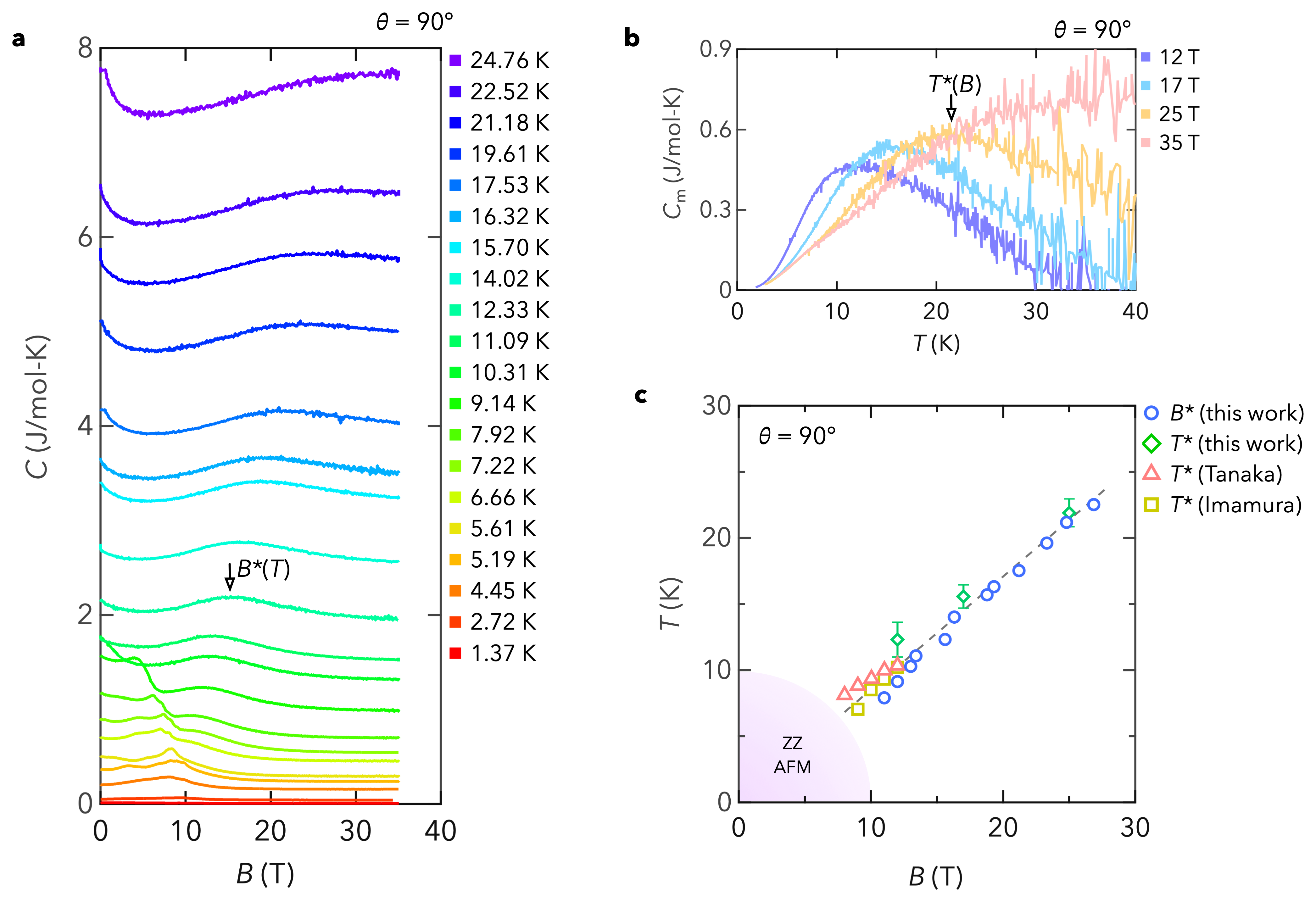} 
\caption{%
{\bf Magnetic field and temperature dependence of specific heat and crossover along the $ab$-plane.}
{\bf a.} Magnetic field dependence of the specific heat for temperatures from 1.37~K to $\sim$25~K along the $ab$-plane. The arrow indicates the crossover field $B^\star(T)$ for 12.33~K. 
{\bf b.} Temperature dependence of specific heat after subtracting the phonon contribution taken from Ref.~\cite{Widmann2019} for several magnetic fields along the $ab$-plane. The arrow indicates the crossover temperature $T^\star(B)$ for 25~T.
{\bf c.} The crossover field $B^\star(T)$ from field dependence (panel~a) and the crossover temperature $T^\star(B)$ from temperature dependence (panel~b) on the $B$ vs. $T$ plane, along with the low field crossover temperature $T^\star(B)$ from Ref~\cite{Tanaka2022} ($\triangle$ markers) and Ref~\cite{Imamura2024} ($\square$ markers). The dashed gray line shows linear dependence ${B^*(T) = 1.17\,T}$.
}
\label{fig:1}
\end{figure*}

$\alpha$-RuCl$_3$ has drawn particular attention as a system that may host the quantum spin liquid state proposed by Kitaev~\cite{Kitaev2006, Jackeli2009}. Despite the presence of antiferromagnetic order at low fields, $\alpha$-RuCl$_3$ exhibits strongly correlated magnetic behavior at high fields~\cite{Baek2017, Winter2018, Widmann2019, Modic2021, Li2021, Zhou2023, Imamura2026}. The evidence for Kitaev quantum spin liquid (KQSL) behavior in $\alpha$-RuCl$_3$ has been the reported half-quantized thermal Hall conductance~\cite{Yokoi2021}, which has recently been questioned in light of a phonon Hall effect~\cite{Shragai2026}. Thermodynamic signatures in this regime~\cite{Tanaka2022, Imamura2026} have likewise been interpreted within the Majorana-flux framework of the Kitaev model, and are usually discussed within the same perturbative regime that underlies the thermal Hall interpretation. This raises the question of whether such assumptions remain justified when interpreting the thermodynamics of $\alpha$-RuCl$_3$ in strong magnetic fields, and thus whether the high-field state shows any evidence for Kitaev-like fractionalized excitations. 

The large Kitaev ($K$) exchange interaction in $\alpha$-RuCl$_3$, of the order of 100~kelvin~\cite{Winter2016, Trebst2022, Matsuda2025}, would naively be expected to set the dominant energy scale governing the low-energy response of the system~\cite{Trebst2022, Matsuda2025}. In the pure Kitaev model~\cite{Kitaev2006}, this exchange sets an intrinsic energy scale, the flux (vison) gap ${\Delta_\mathrm{flux} \approx 0.065|K|}$~\cite{Kitaev2006, Knolle2014, Panigrahi2023}, which is expected to appear as a low-temperature feature, distinct from the gapless itinerant Majorana fermions~\cite{Nasu2015, Trebst2022, Panigrahi2023}. In the weak-field limit, i.e., ${B \ll |K|}$, the Majorana-flux interactions can be treated perturbatively~\cite{Kitaev2006}. To leading order, the flux gap is unchanged by the field, while the field opens a Majorana gap $\Delta_\mathrm{M} \propto h_x h_y h_z/|K|^2$ for a generic field direction~\cite{Kitaev2006, Kaib2019}.
This perturbative regime provides a framework for interpreting low-energy thermodynamic features in terms of interacting Majorana fermions and fluxes~\cite{Widmann2019, Motome2020, Tanaka2022, Imamura2026}. Since the interaction between Majorana and flux sector strengthen with the field, this perturbative treatment is expected to break down at higher fields, although the field scale at which this occurs remains debated~\cite{Kaib2019, Hickey2019, Imamura2026}. At sufficiently high fields, a Zeeman-like gap that is linear-in-$B$ is eventually expected~\cite{Hickey2019, Kaib2019, Yoshitake2020, Li2021}.

Recent magnetotropic measurements~\cite{Modic2021} and quantum Monte Carlo studies~\cite{Sato2024, Inacio2026} find that the magnetic anisotropy response exhibits scaling in $B/T$ over a broad range of field and temperature in the $ab$-plane. Such scaling could means that no intrinsic energy scale, one set by the exchange interactions such as flux gap $\Delta_\mathrm{flux}$ and the field-induced Majorana gap $\Delta_\mathrm{m}$ in the perturbative Majorana-flux regime, enters the thermodynamic response. This raises the question of whether the low-energy thermodynamics retains any signature of a distinct perturbative Majorana-flux regime, or whether a single effective description holds across the field range. However, the magnetotropic coefficient probes the magnetic anisotropy~\cite{Shekhter2023} and does not by itself yield an absolute energy scale or entropy, both of which require a more direct thermodynamic probe.

Here, we report a low-temperature specific heat study of $\alpha$-RuCl$_3$ in high magnetic fields up to 35~T. We find that the specific heat is controlled almost entirely by the effective Zeeman-like energy and its field dependence is different from the expected perturbative Majorana-flux picture. The specific heat thus exhibits a universal $B/T$ scaling for magnetic fields up to the highest fields of 35~T as long as the field is not close to the $c$-axis. Rather than interpreting these observations within a Majorana-flux decomposition motivated by the perturbative Kitaev limit, we suggest that they reflect collective low-energy excitations governed by a single $B$-linear scale, a behavior that may not be specific to $\alpha$-RuCl$_3$ but common to frustrated quantum magnets at high fields. The angular dependence of the specific heat further reveals a pronounced effective $g$-factor anisotropy, characterizing the low-entropy sector, distinct from the $g$-factor of the microscopic ${J_\mathrm{eff}=1/2}$ pseudospin, which is the typical starting point for describing the spins in $\alpha$-RuCl$_3$.

\section{Results}

\begin{figure*}[t!!]
\includegraphics[width=0.72\textwidth]{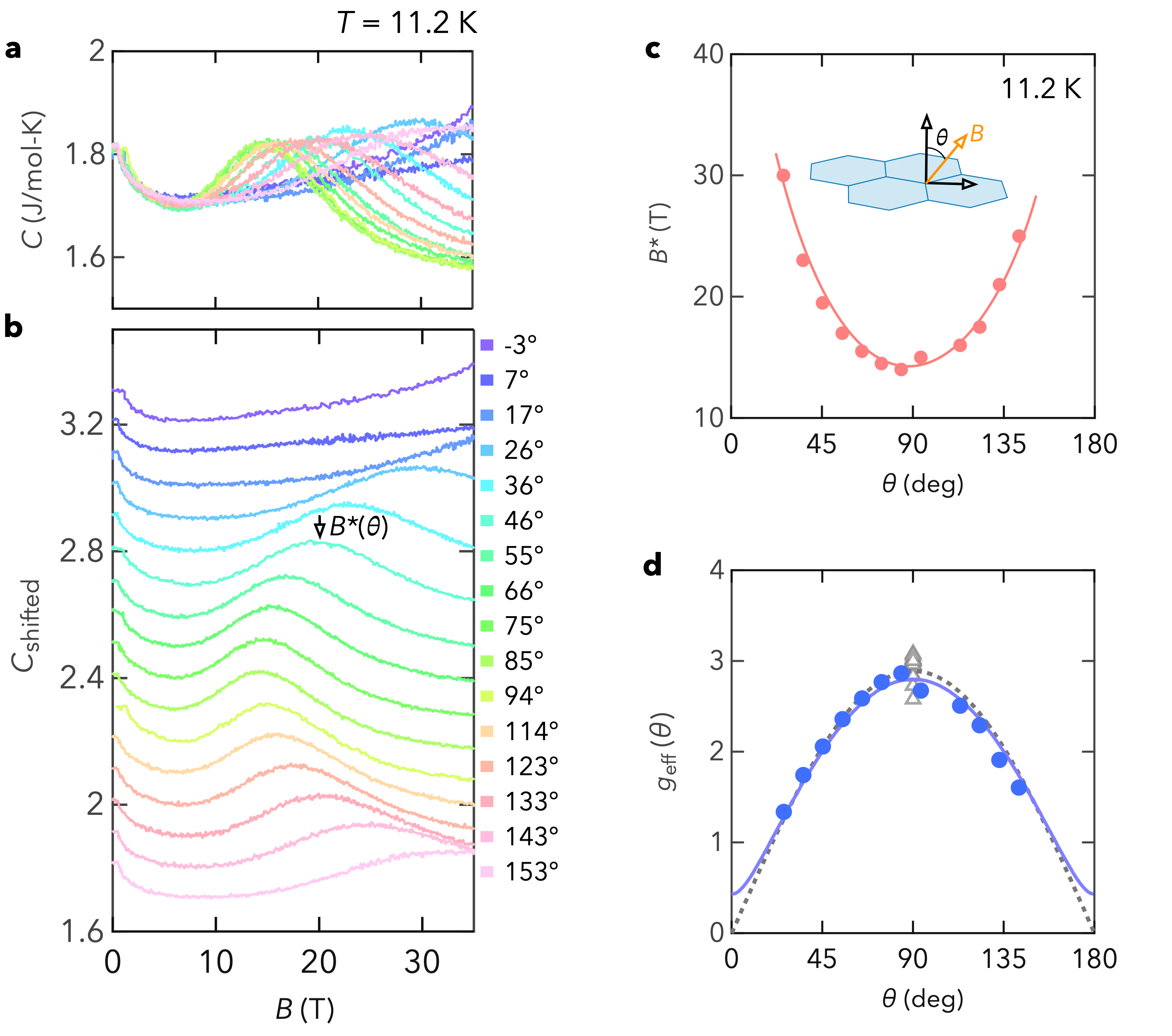} 
\caption{
{\bf Angular dependence of crossover field and determination of the effective $g$-factor.}
{\bf a.} Magnetic field dependence of the specific heat at 11.2~K, slightly above the AFM transition, for different field orientations.
{\bf b.} The field sweeps in panel~a offset vertically for clarity, which identifies the crossover field $B^\star(T,\theta)$ (arrow).
{\bf c.} Angular dependence of the crossover field $B^\star(T,\theta)$ obtained from panels~a,b. Inset shows the field-angle convention, measured from the $c$-axis, so that $\theta=90^\circ$ corresponds to a field in the honeycomb plane. For $\theta\lesssim 25^\circ$ no maximum is resolved up to 35~T.
{\bf d.} Angular dependence of the effective $g$-factor determined from the relation ${\Delta_\mathrm{eff}/k_\mathrm{B}T \simeq 2.4}$, valid for any degenerate two-level system with an energy gap $\Delta_\mathrm{eff}$, ${g_\mathrm{eff}(\theta) = \Delta_\mathrm{eff}/\mu_\mathrm{B} B^\star(T,\theta)}$, where $B^\star(T,\theta)$ is the peak in the field dependence of the specific heat in panel~a at 11.2~K. The blue solid curve is the best fit ${g_{\mathrm{eff}}(\theta)=( g_{\mathrm{eff},c}^2 \cos^2\theta + g_{\mathrm{eff},ab}^2 \sin^2\theta )^{1/2}}$, with $g_{\mathrm{eff},ab}\approx2.8$ and $g_{\mathrm{eff},c}\approx0.4$. The open triangles are obtained from the in-plane crossover field $B^\star(T)$ in Fig.~\ref{fig:1}, and the gray dotted line is the best fit including these points.}
\label{fig:2}
\end{figure*}

In $\alpha$-RuCl$_3$, the magnetic entropy is separated into distinct temperature regimes~\cite{Widmann2019}. The low-temperature entropy, released up to approximately 15\,K, amounts to only $\sim10\%$ of the $R\ln2$ expected for a $J_\mathrm{eff}=1/2$ doublet. A further $\sim50\%$ of $R\ln2$ is released at the scale of the exchange interactions ($\sim$150\,K), and the remaining is recovered only at even higher temperatures~\cite{Widmann2019}. This separation of entropy has been interpreted as thermodynamic evidence for fractionalized excitations, in which the high-temperature entropy is carried by itinerant Majorana fermions at the exchange scale, while the low-temperature entropy is associated with the gapped flux sector~\cite{Widmann2019, Do2017, Nasu2015}. Here we focus on this low-energy sector and study how the corresponding low-energy thermodynamics evolves with magnetic field.

To probe the low-energy excitations, we measure the specific heat as a function of field at fixed temperature. Since the phonon contribution depends strongly on temperature but only weakly on magnetic field, the field dependence of the specific heat at fixed temperature arises predominantly from the magnetic contribution of the sample. We observe a broad feature in the  field dependence of specific heat right above the magnetic transition. Figure~\ref{fig:1}a shows the specific heat $C$ as a function of field for temperatures ranging from 1.37\,K to 24.8\,K. At each temperature above the antiferromagnetic (AFM) phase, this feature exhibits a well-defined maximum, denoted as $B^\star(T)$. For the temperature dependence of specific heat, we use the phonon specific heat obtained by Ref.~\cite{Widmann2019} for $\alpha$-RuCl$_3$. The temperature dependence of the phonon-subtracted specific heat $C_m$ at fixed field exhibits a similar broad maximum denoted as $T^\star(B)$ in Fig.~\ref{fig:1}b. The two determinations of the maximum, obtained independently from field and temperature dependence of specific heat, are mutually consistent (Fig.~\ref{fig:1}c). Combined with the low-field crossover temperatures reported in Refs.~\cite{Tanaka2022, Imamura2024}, $T^\star(B)$ and $B^\star(T)$ together trace out a single, approximately linear relation, $B^\star(T) = \alpha\,T$ where $\alpha = 1.17\,$T/K, shown as dashed line in Fig.~\ref{fig:1}c, extending continuously from the AFM phase boundary up to the highest fields.

\begin{figure*}[t!!]
\includegraphics[width=0.8\textwidth]{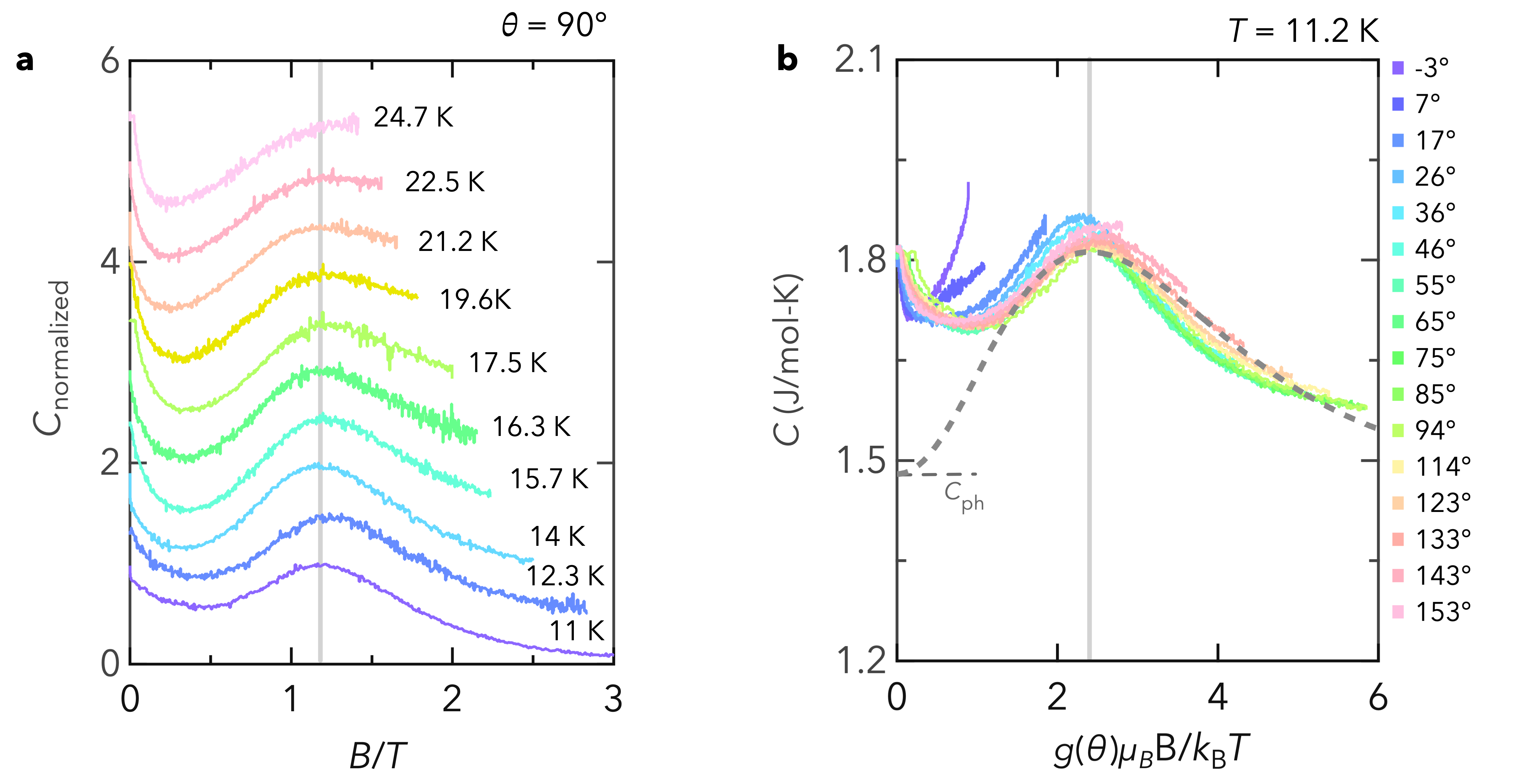} 
\caption{	%
{\bf $B/T$ scaling of specific heat in high fields}  
{\bf a.} $B/T$ scaling of the normalized specific heat at different temperatures in the $ab$-plane. The vertical gray line represents the value $B/T = 1.17$ obtained from Fig.~\ref{fig:1}. 
{\bf b.} Specific heat as a function of $g(\theta)\mu_\mathrm{B}B/k_\mathrm{B}T$ for field dependence data in Fig.~\ref{fig:2}a. The dotted curve represents the functional form of specific heat ${C(B,T)=A(T)\,C_\mathrm{sch}(B,T)+C_\mathrm{ph}(T)}$, where $C_\mathrm{sch}$ is the Schottky function for two-level system and $C_\mathrm{ph}(T)$ is the field-independent phonon term. The vertical gray line represents the peak location condition for a two-level system.  
}
\label{fig:3}
\end{figure*}

At lower fields, this broad feature has previously been interpreted within a perturbative Kitaev description assumed to be valid in that regime~\cite{Tanaka2022, Imamura2024, Imamura2026}. In this description, the position of this broad maximum is associated with the flux sector, ${T^\star(B)\propto\Delta_\mathrm{flux}}$~\cite{Nasu2015, Motome2020, Imamura2024}, and the specific heat is interpreted as the sum of the separate contributions from itinerant Majorana fermions and flux excitations, whose combined form gives rise to the single broad feature observed. Here, we observe that this broad feature in specific heat remains qualitatively similar up to the highest measured fields, well beyond where such a perturbative treatment is expected to remain valid~\cite{Kaib2019, Hickey2019, Patel2019, Gordon2019, Zhang2022}. The associated peak of this feature shifts approximately linearly with magnetic field ${B^\star(T)\propto T}$.  We therefore characterize it using the simpler, minimal Schottky (two-level) description throughout our full field range, and return to what this effective Schottky might represent in the discussion.

A Schottky anomaly in specific heat is the characteristic thermodynamic signature of a finite set of thermally populated energy levels, producing a broad hump in the specific heat wherever the thermal energy $k_\mathrm{B}T$ becomes comparable to the energy level spacing $\Delta$. Many systems host an intrinsic energy gap, such as a crystal-field splitting of the multiplet, which alone produces this hump in the temperature dependence of the specific heat. If the magnetic field additionally couples to the degrees of freedom responsible for this gap, then the gap itself acquires a field dependence, and a corresponding hump appears in the field dependence at fixed temperature. A Zeeman coupling, in particular, produces a gap that depends linearly on magnetic field. In this sense, the peak identified at $B^\star(T)$ provides a direct measure of the characteristic low-energy scale of the system, without requiring an assumed functional form for the underlying density of states. 

We parameterize the observed crossover using the simplest Schottky relation of a two-level system, valid for any doublet with an energy gap, regardless of its microscopic origin. This relation defines an effective energy gap $\Delta_{\mathrm{eff}} \propto B^\star(T)$ directly from the measured crossover peak in the field dependence in Fig.~\ref{fig:1}a. 
To further characterize this effective energy gap, we investigate the angular dependence of the crossover field, rotating the applied field from the $ab$-plane (${\theta = 90^\circ}$) toward the $c$-axis (${\theta = 0^\circ}$). 

Figure~\ref{fig:2}a shows the field dependence of the specific heat at 11.2~K, right above the AFM transition, for several field angles. Vertically shifted field dependence curves shown in Fig.~\ref{fig:2}b for clarity. The phonon contribution is generally field-independent, and thus does not affect the qualitative field-dependence of the specific heat. The resulting angular dependence of $B^\star(T,\theta)$ is shown in Fig.~\ref{fig:2}c.

\begin{figure*}[t!!]
\includegraphics[width=0.78\textwidth]{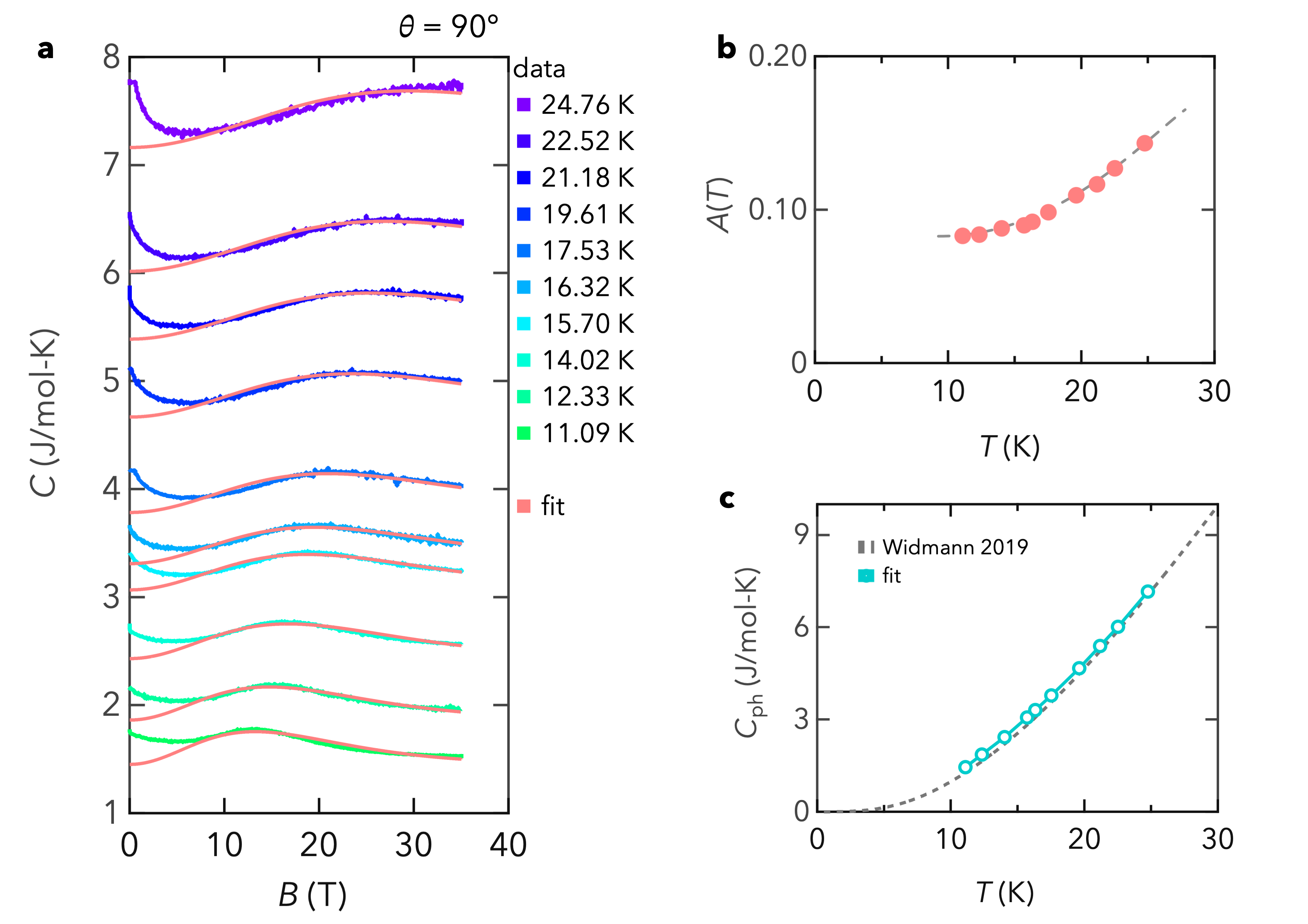} 
\caption{
{\bf Fit and parameters for the field-dependent specific heat in the $ab$-plane.} {\bf a.} Red curves represent the $C(B,T)=A(T)\,C_\mathrm{sch}(B,T)+C_\mathrm{phonon}(T)$, where $C_\mathrm{sch}$ is the two-level Schottky function and $A(T)$ is a temperature-dependent amplitude. The data follow the Schottky form at higher fields and deviate from it at low fields. {\bf b.} Amplitude $A(T)$ obtained from the fits in panel~a. The dashed line is a guide to the eye. {\bf c.} Slope of the low-field upturn as a function of temperature. The dashed line shows the phonon contribution for $\alpha$-RuCl$_3$ from Ref.~\cite{Widmann2019}.
}
\label{fig:4}
\end{figure*}

\begin{figure}[t!!]
\includegraphics[width=1\columnwidth]{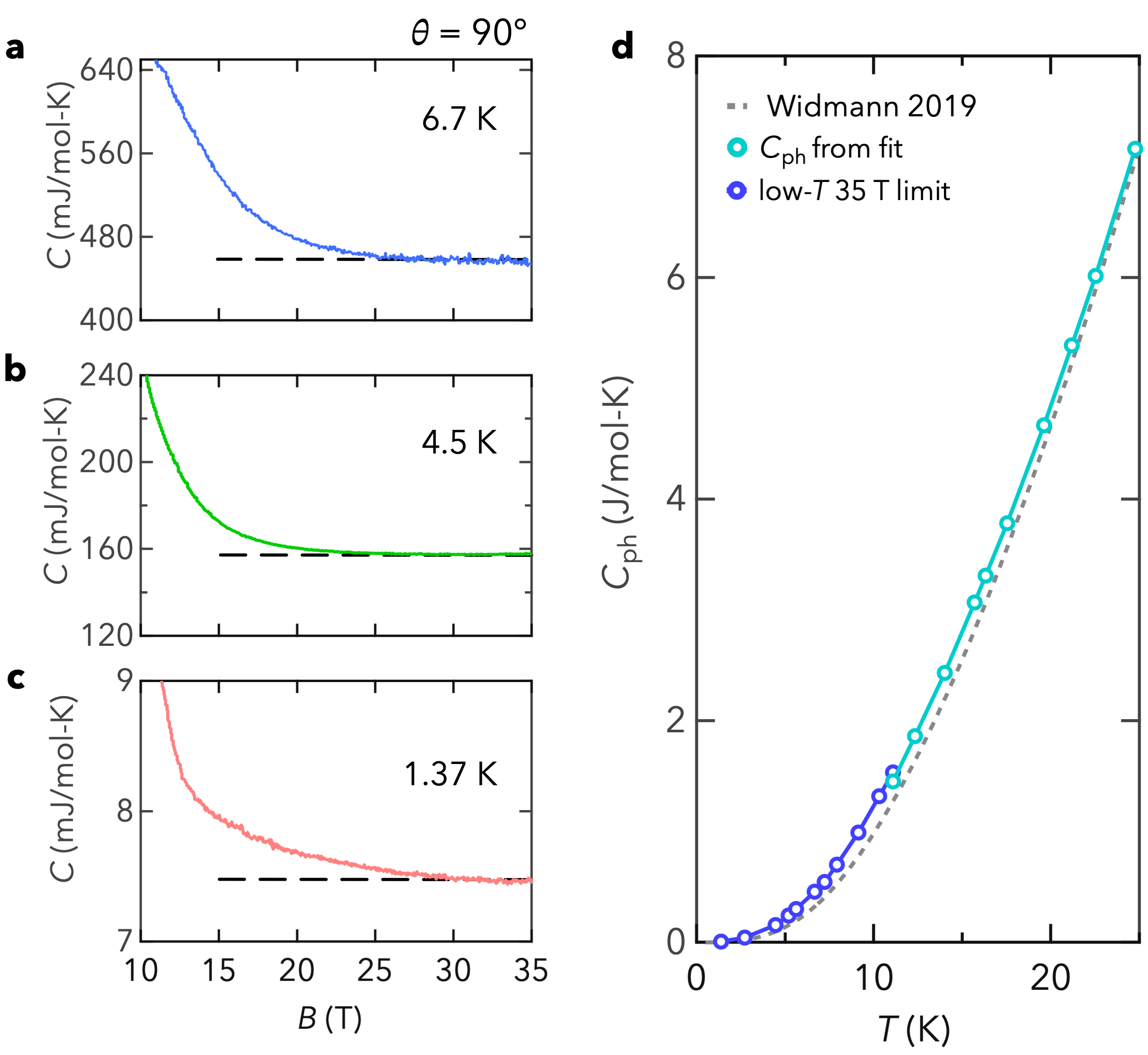} 
\caption{
\textbf{Field dependence of the specific heat at low temperatures and the phonon contribution.}
\textbf{a--c.}~Specific heat $C$ as a function of magnetic field $B$ applied in the $ab$-plane ($\theta = 90^\circ$) at $T = 6.7$, $4.5$ and $1.37$~K. $C(B)$ decreases with increasing field and approaches a field-independent value at the highest fields. The dashed horizontal lines mark this high-field limit.
\textbf{d.}~Temperature dependence of the phonon contribution $C_{\mathrm{ph}}$. Cyan circles are obtained from the Schottky fits over $11$--$25$~K in Fig.~\ref{fig:4}, and blue circles are the high-field limit of the data at $35$~T below $11$~K. The dashed gray line shows the phonon contribution of $\alpha$-RuCl$_3$ from Ref.~\cite{Widmann2019}.
}
\label{fig:5}
\end{figure}

Using the two-level Schottky peak-location condition (${\Delta_\mathrm{eff}(B^\star)/k_BT \simeq 2.4}$), we extract an effective $g$-factor, ${g_\mathrm{eff}(\theta) = \Delta_\mathrm{eff}(B^\star)/\mu_\mathrm{B}B^\star(T,\theta)}$, associated with this low-energy excitation (Fig.~\ref{fig:2}d). The resulting angular dependence is well described by ${g_{\mathrm{eff}}(\theta) = (g_{\mathrm{eff},c}^2\cos^2\theta + g_{\mathrm{eff},ab}^2\sin^2\theta)^{1/2}}$, with $g_{\mathrm{eff},ab}\approx2.8$ and $g_{\mathrm{eff},c}\approx0.4$, revealing a strongly anisotropic effective $g$-factor, with a ratio $g_{\mathrm{eff},ab}/g_{\mathrm{eff},c}\approx7$. We emphasize that $g_{\mathrm{eff}}(\theta)$ is an effective parameter that characterizes the low-energy excitation as measured, and need {\it not} correspond to the bare $g$-factor of the microscopic pseudospin $J_\mathrm{eff}=1/2$ doublet; rather, $g_{\mathrm{eff}}(\theta)$ encapsulates the combined effect of exchange interactions and the microscopic $g$-tensor on the observed low-energy response. These excitations are emergent in the sense that they carry very small entropy compared to the microscopic $J_\mathrm{eff}=1/2$ pseudospin and couple to the field with an effective $g$-tensor that reflects combined effect of both the exchange interactions and the microscopic $g$-tensor of the pseudospin.

The strongly anisotropic but $B$-linear form of $\Delta_\mathrm{eff}$ suggests that this Zeeman-like coupling alone sets the low-energy scale of the system throughout the accessible field-temperature range. If so, the specific heat should depend on field and temperature only through the ratio $B/T$. Testing this over the full field range, and not only at its maximum, probes whether any additional energy scale, such as one set by the exchange interactions and expected in a perturbative Kitaev description, enters the low-energy thermodynamics. Figure~\ref{fig:3}a shows specific heat for the $ab$-plane, normalized and offset vertically for clarity, plotted against $B/T$ for temperatures above the antiferromagnetic transition. The characteristic crossover feature identified in Fig.~\ref{fig:1} occurs at approximately the same value of the ratio $B/T \approx 1.17$\,T/K across this entire temperature range. 

To look at the dominant contribution in the specific heat, we use the effective field-angle dependent energy gap ${\Delta_{\mathrm{eff}}(B,\theta) = g_\mathrm{eff}(\theta)\,\mu_\mathrm{B}\,B}$, where the effective $g$-factors were extracted from the peak location in Fig.~\ref{fig:2}c. The specific heat as a function of $g(\theta)\mu_B B/k_\mathrm{B}T$ for different field orientations at 11.2~K is shown in Fig.~\ref{fig:3}b. The dashed line represents ${C(B,T) = A(T)\,C_\mathrm{sch}(B,T) + C_{\mathrm{ph}}(T)}$, where $C_\mathrm{sch}(B,T)$ is the standard two-level Schottky function, and the amplitude $A(T)$ is the number of two-level systems per formula unit, so that $A\,R\ln 2$ is the maximum entropy that can be released. The phonon contribution $C_{\mathrm{ph}}(T)$ is assumed to be independent of magnetic field at each temperature. The scaling works well across all measured fields except at low fields where an upturn is observed. This effective description seems to be valid for all orientations except close to the $c$-axis ($\theta \sim 0^\circ$), suggesting that the effective description may not hold for fields along the $c$-axis.

At low fields the specific heat departs from the Schottky form: $C$ rises steeply as $B\to0$ (Figs.~\ref{fig:2},\ref{fig:3}). This upturn could be an additional contribution that is present at zero field and quickly gets suppressed by modest fields for all temperatures. It is consistent with the large zero-field specific heat above the ordering temperature, which the field rapidly removes. To quantify the Schottky contribution above this low-field region, we extract the temperature dependence of the amplitude $A(T)$ and of the phonon $C_\mathrm{ph}(T)$ by fitting the field dependence with the functional form of $C(B,T)$ described earlier. The Schottky contribution takes the standard two-level form,
\begin{align}
    \frac{C_\mathrm{sch}(x)}{R}=  x^2 \frac{e^{x}}{(1+e^{x})^2},
\end{align}
where $x = g\mu_\mathrm{B}B/k_\mathrm{B}T$, and $R$ is the universal gas constant. The field dependence of specific heat at different temperatures in the $ab$-plane along with the fit (red curves) are shown in Fig.~\ref{fig:4}a.

The amplitude $A(T)$ increases from about 0.08 at 11~K to about 0.14 at 25~K (Fig.~\ref{fig:4}b). A closed two-level system would have a temperature-independent $A$. Its growth indicates that the low-energy sector is not a closed doublet and that additional states become thermally accessible with increasing temperature, consistent with most of the entropy of the $J_\mathrm{eff}=1/2$ doublet being released at higher temperatures~\cite{Widmann2019}. Since $A$ cannot exceed one doublet per Ru$^{3+}$, this growth must saturate at higher temperatures. The phonon $C_\mathrm{ph}(T)$ rises from about 1.4 to 7.2~J/mol-K over the same range and follows the lattice contribution of Ref.~\cite{Widmann2019} closely; however, it is slightly higher at low temperatures (Fig.~\ref{fig:4}c).

For low temperatures, the specific heat decreases monotonically with field beyond the antiferromagnetic critical field and approaches a nearly field-independent value at 35~T (Fig.~\ref{fig:5}a--c). Figure~\ref{fig:5}d shows the high field saturation value of specific heat below 11~K as blue markers. The 35~T limit saturation value continues smoothly and coincides with the $C_\mathrm{ph}(T)$ values obtained by the fit in Fig.~\ref{fig:4}. Both follow a single curve that lies slightly above the lattice contribution from Ref.~\cite{Widmann2019} (shown as a dotted curve in Fig.~\ref{fig:5}d) at low temperatures.

\section{Discussion}
The additional exchange interactions such as Heisenberg and off-diagonal terms, which are known to be present in $\alpha$-RuCl$_3$, give rise to the magnetic ordering in the system. These additional terms would generically introduce behavior that is not completely accounted for by the pure Kitaev model which does not contain these additional interactions. An applied magnetic field further modifies the behavior of the system. 

It is worth asking whether, in $\alpha$-RuCl$_3$, there is a magnetic field regime that plausibly falls within a perturbative Kitaev description in the first place. Theoretical estimates place the perturbative window at only a small fraction of the exchange scale, with $B\approx 0.03|K|$ for the ferromagnetic Kitaev limit believed relevant to $\alpha$-RuCl$_3$~\cite{Kaib2019, Janssen2019}. This corresponds to a field of only a few Tesla for $K$ of order 100~kelvin, well below the range over which we observe the broad feature in specific heat and its associated $B/T$ scaling. Consistent with this estimate, we see deviations from the Schottky-like description at low fields, the narrow field regime below 5~T in which the perturbative treatment may be expected to apply.

A more complete comparison is the non-perturbative density-matrix renormalization group (DMRG)~\cite{Li2021} and exact diagonalization (ED)~\cite{Gordon2019} calculations that include such additional interactions. For magnetic fields in $ab$-plane, the DMRG calculations of specific heat find a transition from an ordered state to a field-polarized state, with a crossover energy scale that vanishes at the critical field and then increases linearly, nearly twice as steeply as the single scale we observe in measurements~\cite{Li2021}. For fields in the $c$-axis, both the DMRG and ED calculations predict an intermediate quantum spin liquid ground state before field polarizes the spins~\cite{Gordon2019, Li2021}. 

At high enough fields, the system is expected to approach a field-polarized state, in which the lowest excitations acquire a Zeeman-like gap that grows linearly with field~\cite{Kaib2019, Li2021}. The $B$-linear scale we observe is consistent with such a state. However calculations find that this scale does not extrapolate linearly to zero at zero field but is affected by the proximity of the antiferromagnetic phase at low fields~\cite{Li2021, Sato2024, Inacio2026}, which would generically produce a departure from $B/T$ scaling when fields are closer to the AFM phase boundary. By contrast, we observe a single relation ${B^\star(T)\propto T}$ that extends continuously down to the antiferromagnetic phase boundary. 

Such behavior in magnetic field is not unique to $\alpha$-RuCl$_3$. Similar behavior governed by a single ratio $B/T$ rather than by any intrinsic scale, has been reported in a range of frustrated quantum magnets with distinct microscopic descriptions, including the honeycomb iridate H$_3$LiIr$_2$O$_6$~\cite{Kitagawa2018}, the molecular cluster magnet LiZn$_2$Mo$_3$O$_8$~\cite{Sheckelton2014}, the kagome antiferromagnet herbertsmithite~\cite{Murayama2022}, and the triangular-lattice charge-density-wave compound 1T-TaS$_2$~\cite{Klanjsek2017}. This diversity has been proposed to reflect a disordered, random-singlet-like state, in which quenched disorder produces a broad spread of local energy scales~\cite{BhattLee1982, Fisher1994, Kimchi2018, Kimchi2018b}. 

We emphasize that the observed $B/T$ scaling alone does not imply quantum criticality. A simple non-interacting multiplet with Zeeman-split levels also satisfies $C=f(B/T)$, yet it lacks the scale-invariant power-law asymptotics that characterize a quantum critical regime. Our data show no such power-law behavior or divergence in either the low-field or the high-field limit of the scaling function, consistent with the conclusion of quantum Monte Carlo simulations of the extended Kitaev model for $\alpha$-RuCl$_3$~\cite{Sato2024}. Our results suggest that, for fields away from the $c$ axis, the high-field thermodynamics of $\alpha$-RuCl$_3$ is governed by collective low-energy excitations with a single effective $B$-linear energy scale, and that no additional scale set by the exchange interactions is resolved within the measured range of field and temperature. For fields along the $c$-axis, the extrapolated small effective $g$-factor in that direction pushes the corresponding crossover beyond the maximum field of our measurements, so characterizing the high-field state in this direction will require measurements at higher fields.

\section{acknowledgments}
\begin{acknowledgments}
We thank Marcus Schmidt and Michael Baenitz for providing single crystals. Resistive high field 35~T magnet measurements were performed at the National High Magnetic Field Laboratory (NHMFL), which is supported by the National Science Foundation Cooperative Agreement No. DMR-1157490 and the State of Florida. K.A.M. and A.K. acknowledge support from the European Research Council Starting Grant 101078696-TROPIC. F.A. and K.A.M. acknowledge financial support from the Austrian Science Fund (SFB F 86, Q-M\&S). A.R. acknowledges support by the Swedish Research Council, Grant No. 2021-04360.
\end{acknowledgments}

\bibliographystyle{apsrev4-1}
\bibliography{RuCl3-arXiv.bib}

\end{document}